\documentclass{article}
\usepackage[utf8]{inputenc}
\usepackage{hyperref}
\usepackage{times}
\usepackage{graphicx}
\usepackage{amssymb}
\usepackage{amsthm}
\usepackage{amsmath}
\usepackage{dsfont}
\usepackage{bm}
\usepackage{mathrsfs}
\usepackage{bbold}
\usepackage{gensymb}
\usepackage{doi}

\title{Polariton fluids for analogue gravity physics}
\author{M. J. Jacquet \and T. Boulier \and F. Claude \and A. Ma\^itre \and E. Cancellieri \and C. Adrados \and A. Amo \and S. Pigeon \and Q. Glorieux \and A. Bramati \and E. Giacobino}

\begin{document}

\maketitle
\begin{abstract}
Analogue gravity enables the study of fields on curved spacetimes in the laboratory. There are numerous experimental platforms in which amplification at the event horizon or the ergoregion has been observed. Here, we demonstrate how optically generating a defect in a polariton microcavity enables the creation of one- and two-dimensional, transsonic fluid flows. We show that this highly tuneable method permits the creation of sonic horizons. Furthermore, we present a rotating geometry akin to the water-wave bathtub vortex. These experiments usher-in the possibility of observing stimulated as well as spontaneous amplification by the Hawking, Penrose and Zeld'ovich effects in fluids of light.
\end{abstract}
\section{Introduction}
Analogue gravity experiments use well-controlled laboratory systems to study phenomena that are either currently eluding our comprehension or simply unreachable by their very nature, such as the behaviour of small amplitude waves near black holes  \cite{unruh_experimental_1981}.
For example, it is possible to create static black- \cite{philbin_fiber-optical_2008,lahav_realization_2010,choudhary_efficient_2012,euve_scattering_2018,Jacquet_book_2018,munoz_de_nova_observation_2019} or white-hole \cite{philbin_fiber-optical_2008,rousseaux_observation_2008,choudhary_efficient_2012,faccio_analogue_2010,weinfurtner_measurement_2011,nguyen_acoustic_2015,Rousseaux_PRL_2016,Jacquet_book_2018}  event horizons for waves in media and rotating geometries with horizons as well as ergo-regions \cite{torres_rotational_2017,torres_analogue_2019}.
Thus, analogue gravity enables the study of classical and semi-classical effects of fields on curved spacetimes, such as the Hawking effect at static horizons \cite{hawking_black_1974,hawking_particle_1975} or superradiance \cite{zeldovich__1970,penrose_extraction_1971,zeldovich_amplification_1972,Solnyshkov_penrose_2019} and the quasi-normal modes of rotating geometries \cite{patrick_black_2018}.
These effects were recently observed in water experiments \cite{euve_scattering_2018,torres_rotational_2017,torres_analogue_2019}.
In these seminal experiments, excitations were seeded by a classical state to stimulate the emission. Unfortunately, because of its low temperature, spontaneous emission (stemming from quantum fluctuations at the horizon or the ergo-surface) cannot be observed in these systems. This can be done with fluids at lower temperatures, such as atomic BECs \cite{munoz_de_nova_observation_2019}.
Here we discuss another experimental platform for analogue gravity: polariton fluids in microcavities.

Fluids of light can display properties similar to those of a matter fluid when the environment such as confinement in an optical cavity provides the photon with an effective mass and when the optical nonlinearity of the medium provides effective photon-photon interactions.
With polaritons in microcavities, two-dimensional \footnote{Unless specified otherwise, we write one- and two-dimensional in reference to the spatial dimensions of the system. There is always one temporal dimension in addition.} fluids in the plane of the cavity have been demonstrated to exhibit Bose-Einstein condensation \cite{Kasprzak_BECpolariton_2006,Balili_BECpolariton_2007} and superfluidity \cite{Lagoudakis_fluidlightexp_2008,Utsunomiya_fluidlightexp_2008,Amobloch_fluidlightexp_2009,Amo_fluidlightexp_2009}.
These hydrodynamic properties, together with the wide range of available control and engineered tunability of the optical setups, enable the simulation of systems described by Hamiltonians including various kinds of interactions.
For example, elementary excitations of a polariton fluid can experience an effective curved spacetime determined by the physical properties of the flow \cite{marino_acoustic_2008}.

A flow featuring a horizon was first realised with a  polariton fluid by means of an engineered defect in the cavity \cite{nguyen_acoustic_2015}.
Here, in contrast, we show how an optical defect can be used to realise a transsonic flow.
This technique has the advantage to be widely adaptable to various cavity geometries without requiring the mechanical manufacturing of a new geometry for each gravity scenario to be investigated.
We demonstrate this advantage by realising three scenarios: a one-dimensional and a two-dimensional static horizons and a draining, rotating vortex flow. 

The paper is structured as follows: we first introduce the quantum fluid properties of microcavity polaritons.
We show how sub- and supersonic flows may be created by propagation across an optical defect and discuss the dispersion relation of Bogoliubov modes in these flows.
We then present an experiment demonstrating the advent of superfluidity in a two-dimensional flow.
In a second part, we revisit the canonical derivation of the wave equation for these modes and discuss the analogy with a curved spacetime metric.
In the fourth section, we present experimental data demonstrating the creation of one- and two-dimensional static horizons for the Bogoliubov modes before moving to a rotating geometry in the final section.

\section{Polariton fluids in microcavities}
\begin{figure*}[ht]
    \centering
    \includegraphics[width=.95\textwidth]{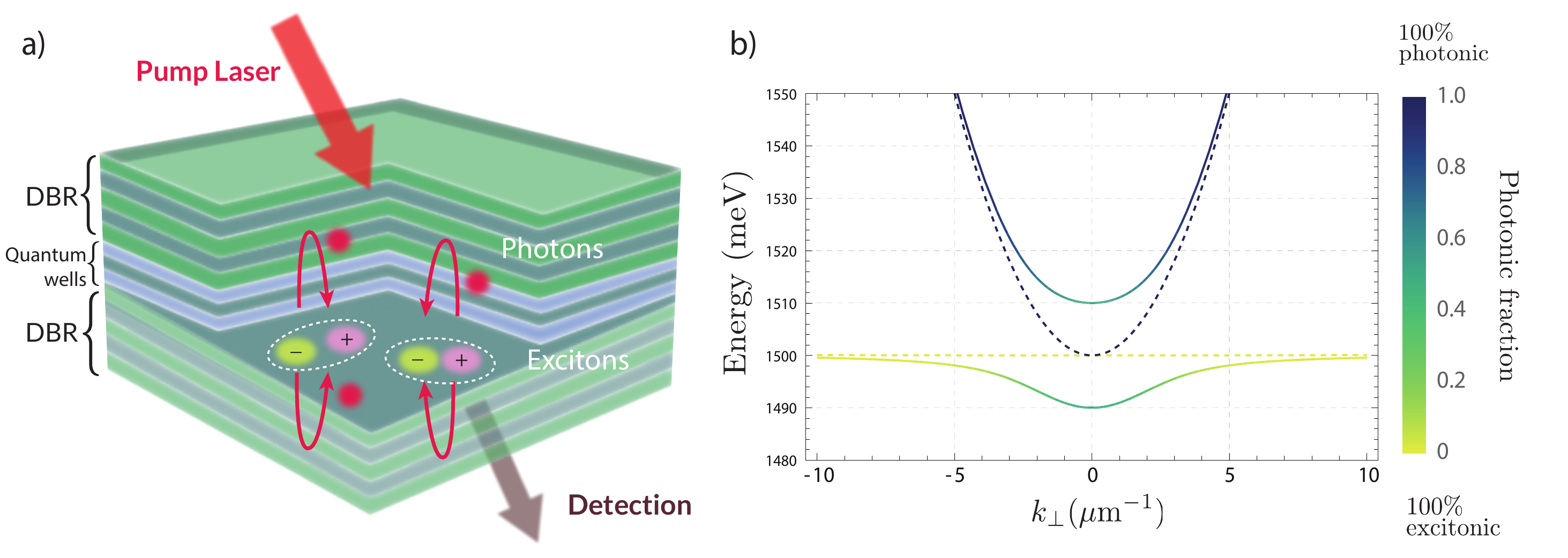}
    \caption{Microcavity polaritons. \textbf{a)} Scheme of the semiconductor microcavity, with InGaAs quantum wells between Bragg mirrors. \textbf{b)} Dispersion curve of the strongly coupled excitons and cavity photons, giving rise to upper (UP) and lower (LP) polaritons.\label{fig:setup}}
\end{figure*}

The system used for analogue gravity is a semiconductor microcavity, made of quantum wells of InGaAs located between two Bragg mirrors forming a thin cavity of high finesse, as shown in Fig.\ref{fig:setup} \textbf{a)}.
In the quantum wells, electrons and holes form bound states, named excitons, that couple efficiently to photons.
In these conditions the strong coupling regime can be reached between excitons and photons and the vacuum Rabi splitting takes place \cite{arakawa_polariton_1992}, resulting in mixed photon-exciton states, known as cavity polaritons.
The two eigenstates, the upper and the lower polariton are shown in Fig.\ref{fig:setup} \textbf{b)}.
They are separated by 5meV, and in the following we will work with the lower polariton branch, excited close to resonance by a CW, monochromatic laser.
To avoid thermal noise, the system is operated at 5K in a Helium cryostat.

We are interested in the propagation of light in the cavity plane, with a small incidence angle. In an optical cavity, the resonance condition of an incoming laser with a cavity mode is given by $p\frac{\lambda}{2}=l\cos{\theta'}$ with $p$ an integer (here $p=4$), $\lambda$ the mode wavelength, $l$ the cavity thickness and $\theta'$ the optical angle of incidence (inside the cavity).
This implies that the component of the wavevector $k$ perpendicular to the cavity plane, $k_z$, is fixed to $p\pi/l$.
In the case of a small angle of incidence (or a small component of the wavevector parallel to the cavity plane $k_x$) the dispersion can then be written as
\begin{equation}
\label{eq:disprelpol}
    \omega=c\sqrt{k_z^2+k_x^2}\approx c k_z +\frac{\hbar k_x^2}{2m},
\end{equation}
where $m$ is the effective photon mass for motion in the cavity plane, $m= \hbar k_z/c$ .
This geometry gives a parabolic dispersion curve for the photons inside the cavity.
The strong coupling modifies the photon and the exciton dispersion curves, introducing a curvature on the lower polariton branch, with an effective mass similar to that of the photon. with such a dispersion curve the group velocity of the polaritons can now be written as $v= \hbar k_x/m$.
Moreover, excitons interact together via electromagnetic interaction, resulting in an interaction between polaritons.

As a result, cavity polaritons are interacting composite bosons with a small effective mass ($\sim10^{-5}\mathrm{m_e}$).
They have a large coherence length of the order of 1 to 2 $\mathrm{\mu m}$ at the operating temperature (5K).
This enables the building of many-body coherent quantum effects, such a condensation and superfluidity at these temperatures.
The quantum fluid properties of photons in a cavity and polaritons were predicted \cite{Boyce_fluidlight_1999,carusotto_fluidlightproposal_2004,ciuti_fluidlightproposal_2005} and later observed in several groups \cite{Kasprzak_BECpolariton_2006,Balili_BECpolariton_2007,Lagoudakis_fluidlightexp_2008,Utsunomiya_fluidlightexp_2008,Amo_fluidlightexp_2009,Amobloch_fluidlightexp_2009}.
The main interest of this system for quantum simulation is the evolution of the polariton wavefunction, $\psi$, which is described by an equation similar to the Gross-Pitaevskii equation:
\begin{equation}
    \label{eq:GPE}
    i\hbar\frac{\partial\psi}{\partial t}=-\frac{\hbar^2}{2m}\nabla^2\psi+ g |\psi|^2\psi+V_{ext}\psi-i\gamma\psi+F_p.
\end{equation}
The first term on the right-hand side represents the kinetic energy, the second term is the nonlinear interaction, the third term is an external potential $V_{ext}$, the fourth and fifth terms are the losses ($\gamma$) due to the cavity and the resonant pumping by a laser.
These last two terms are additions to the usual Gross-Pitaevskii equation: they account for the driven-dissipative nature of the system, with losses of light out of the cavity, compensated by pump photons.

\begin{figure*}[ht]
    \centering
    \includegraphics[width=.95\textwidth]{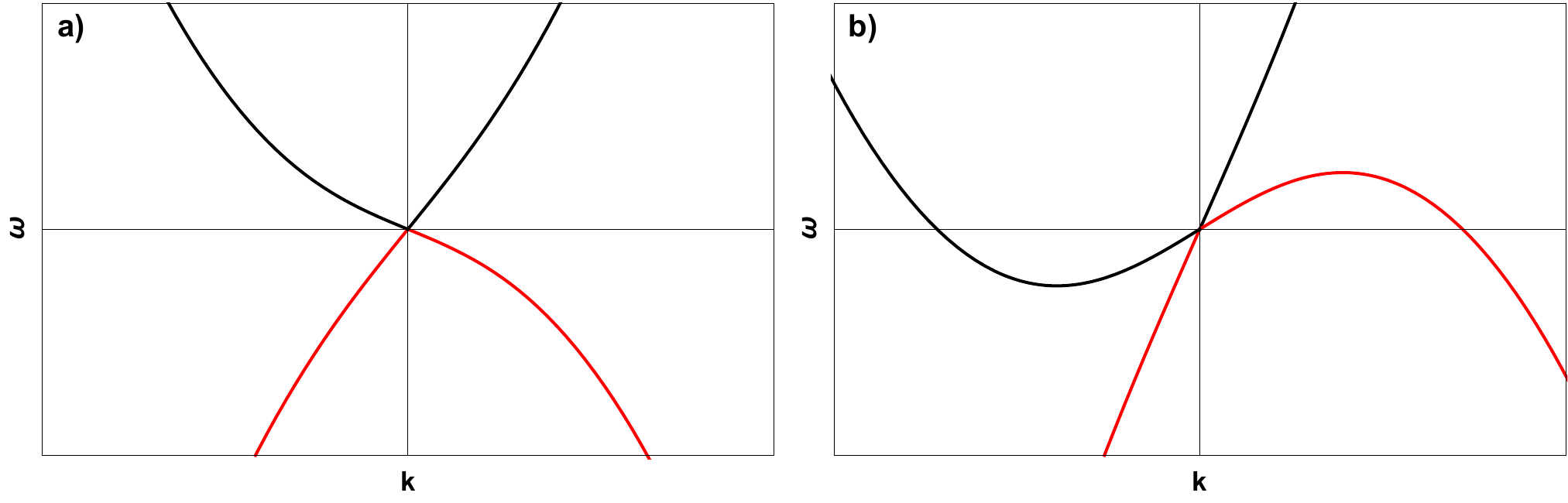}
    \caption{Sonic dispersion relation \eqref{eq:disprelbog} in the laboratory frame: \textbf{a)} subsonic fluid flow (superfluid regime); \textbf{b} supersonic fluid flow. Black --- positive fluid-frame frequency branch; Red --- negative fluid-frame frequency branch.
    \label{fig:disprel}}
\end{figure*}

The onset of superfluidity can be studied using the Boguliubov method, where the effect of a small perturbation is studied by linearizing the Gross-Pitaevskii equation in the vicinity of the operating point.
The resulting Bogoliubov dispersion is given by
\begin{equation}
    \label{eq:disprelbog}
    \omega(k)=\pmb{\delta k.v_0} \pm \sqrt{\frac{\hbar\delta k^2}{2m}\left(\frac{\hbar\delta k^2}{2m}+2 gn\right)},
\end{equation}
where $v_0$ is the velocity of the polariton flow and $\pmb{\delta k=k-k_p}$ is the mean difference between the wavevector of the perturbation and the wavevector of the pump and $n=|\psi_0|^2$ is the polariton density.
Eq\eqref{eq:disprelbog} was established from the Gross-Pitaevskii equation for matter superfluids. For laser driven systems like polaritons, it is only valid if the detuning between the laser energy and the lower polariton energy (in the absence of interaction) is equal to the nonlinear interaction energy, which corresponds to the upper turning point of the bistability curve \cite{ciuti_fluidlightproposal_2005,carusotto_ciuti_RMP_2013} . This is the point where the superfluidity demonstrations in polaritons were performed \cite{Amo_fluidlightexp_2009}.
In the rest frame of the fluid, the full dispersion relation is centro-symmetric around the $\{k=0,\omega=0\}$ point --- there are two branches : a positive (fluid frame) frequency branch and a negative (fluid frame) frequency branch (the so-called "ghost branch"). In Fig.\ref{fig:disprel} we show the dispersion relation in the laboratory frame: it is boosted from the fluid frame by the Doppler effect ($\omega=\omega'- v_0 k$) with a velocity $v_0$ that depends on the flow velocity in each region (subsonic in \textbf{a)}, supersonic in \textbf{b)}). The positive (negative) fluid-frame frequency branch is shown in black (red).
Depending of the value of $\delta k$ compared to the inverse of the healing length $\xi=\sqrt{\hbar/mgn}$ the dispersion curve can be changed considerably.
If $\delta k$ is large, $\delta k \xi\gg1$, the above equation gives $\omega (\delta k)=\pm \frac{\hbar\delta k^2}{2m}+ gn$ --- the usual parabolic dispersion for a massive system, shifted by the interaction energy $2gn$.
If $\delta k$ is small,  $\delta k \xi\ll1$ we have $\omega (\delta k)=\pmb{\delta k.v_0} \pm c_s |\delta k|$  with the speed of sound $c_s=\sqrt{(\hbar gn/m)}$.
Fig.\ref{fig:disprel} shows a sonic dispersion with a discontinuity in the slope at $\delta k=0$.
This point corresponds to the minimum (maximum) of the positive-(negative) frequency branch, as can be seen in Fig.\ref{fig:disprel} \textbf{a)}.
Moreover, in the subsonic case ($v_0 < c_s$), the slopes on each side of $\delta k=0$ have opposite signs.
This gives rise to superfluidity because no elastic scattering (for example against an obstacle) is possible in this case, thus fulfilling the Landau criterion.
The critical velocity below which superfluidity occurs has been shown here to be the sound velocity. However, in some experiments the presence of an incoherent reservoir of excitons modifies the interaction energy and should be taken into account for the definition of the critical velocity \cite{Richard_dispersion_2019}.
In the following $c_s$ will denote this critical velocity.

The experiments demonstrating this effect \footnote{The light transmitted by the cavity is observed in the near field and in the far field.} were performed by shining a laser quasi-resonant with the lower polariton branch on the microcavity with a small angle of incidence \cite{Amo_fluidlightexp_2009}.
This configuration gives a small velocity to the fluid of light in the cavity plane, so that it is smaller than the speed of sound at high intensity.
The dispersion for Bogoliubov excitations is then that of Fig.\ref{fig:disprel} \textbf{a)}.
On the other hand, when the fluid flow is supersonic, the dispersion relation is as in Fig.\ref{fig:disprel} \textbf{b)}.

\begin{figure*}[ht]
    \centering
    \includegraphics[width=.85\textwidth]{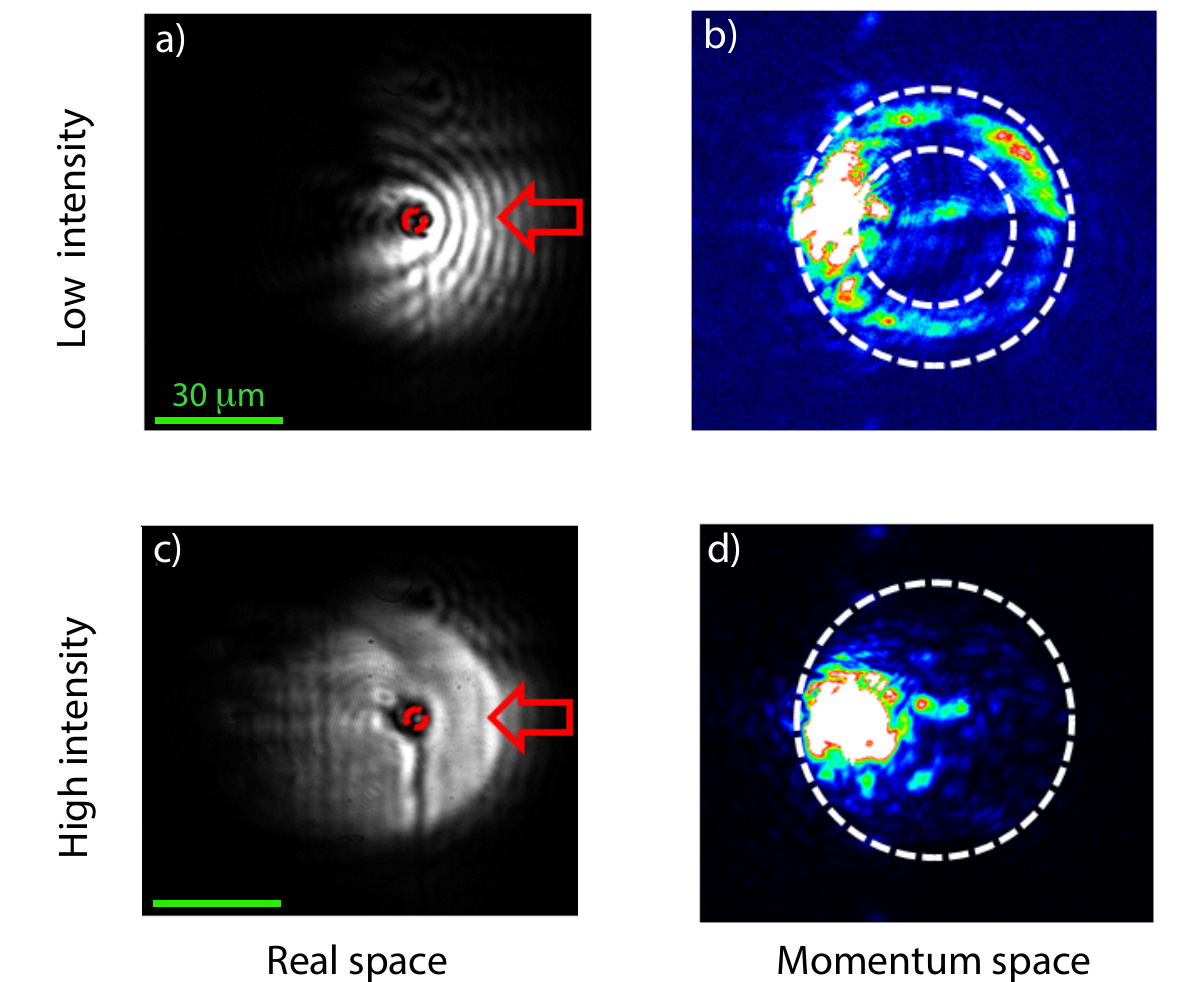}
    \caption{Polariton flow around an optical defect. Left : near-field  images of the polariton. Flow density around a defect for low (upper panel)  and high (lower panel) excitation power. The red arrow shows the direction of the flow towards the defect. At low power the polariton fluid scatters on the defect, giving rise to parabolic wavefronts, while at high power there is no more scattering and the fringes disappear. This is confirmed by the momentum distribution on the far-field images, showing an elastic scattering ring in the low power case, which disappears at high power.\label{fig:1Dflow}}
\end{figure*}

In Fig.\ref{fig:1Dflow}, we show what happens when the fluid of light hits a defect. 
There is a drastic change when, by increasing the polariton density, one goes from the linear regime (where the particle-particle interactions are negligible, top row) to the nonlinear regime (bottom row):
in Fig.\ref{fig:1Dflow} \textbf{c)}, we see interference fringes between the incoming flux and the scattered light disappear in the nonlinear regime.
As can be seen by comparing Figs.\ref{fig:1Dflow} \textbf{b)} and \textbf{d)}, the scattering ring in momentum space also disappears.
This is due to superfluidity.
Importantly, the flow velocity goes from subsonic (on the right hand side of Fig.\ref{fig:1Dflow} \textbf{c)}) to supersonic (on the left hand side of Fig.\ref{fig:1Dflow} \textbf{c)}) over the optical defect.

\section{Analogue gravity with polaritons}\label{sec:anagrav}
The ability to generate a trans-sonic flow in one and two spatial dimensions is at the basis of the use of polariton quantum fluids for analogue gravity.
Here, we revisit the derivation establishing the analogy and explain how polariton flows can be tailored to generate the corresponding experimental configuration.

We begin by converting the Gross-Pitaevskii wave equation for the polaritons into hydrodynamic equations using Madelung transformation. At this point it should be noted that the mathematical analogy \cite{unruh_experimental_1981} between the wave equation for Bogoliubov excitations in a fluid flow and on a curved spacetime is valid at low $k$ only \cite{visser_acoustic_1998}, where the dispersion relation \eqref{eq:disprelbog} is linear --- the excitations behave as genuine sound waves with a constant group velocity $c_s$.
In this `hydrodynamic regime', the influence of the pump, losses and external potential on the kinematics of sound waves is negligible, and thus the last three terms of \eqref{eq:GPE} may be dropped \footnote{When considering excitations at all possible $k$s, the analogy with wave motion on curved spacetimes is more complicated but can still be established on kinematic grounds \cite{jacquet2020influence}.} the wave equation reduces to the nonlinear Schr\"odinger equation (NLSE) \cite{Agrawal_2012}:
\begin{equation}
\label{eq:NLSE}
i\partial_t \psi=-\frac{\hbar}{2m}\nabla^2\psi+g|\psi|^2 \psi
\end{equation}
As in \eqref{eq:GPE}, $\nabla^2\psi$ accounts for the kinetic energy, while the second term describes the nonlinear interaction (self-defocusing), with $|\psi|^2$ the polariton intensity .
We convert Eq.\eqref{eq:NLSE} into the hydrodynamic continuity and Euler equations by writing the complex scalar field in terms of its amplitude and phase --- $\psi=\sqrt{\rho}e^{i\phi}$, with $\rho$ the polariton density, corresponding to the fluid density:
\begin{equation}
\label{eq:hydro}
    \partial_t\rho+\pmb{\nabla}\cdot(\rho \pmb{v})=0,
\end{equation}
\vspace{-.25cm}
\begin{equation}
    \label{eq:eulereq}
    \partial_t\phi+\frac{1}{2\hbar}m \pmb{v}^2+g\rho-\frac{\hbar}{2m}\frac{\Delta\rho^{1/2}}{\rho^{1/2}}=0,
\end{equation}
where $\pmb{v}=\frac{\hbar}{m}\pmb{\nabla}\phi$ is the fluid velocity.
Note that, here, $\pmb{v}$ is a two-dimensional vector ($\pmb{v}=\{v^x,v^y\}$ with $x$ and $y$ the spatial coordinates) and $v$ is its modulus.

We obtain the wave equation for the Bogoliubov excitations (sound waves at low $k$) by linearising eqs. \eqref{eq:hydro}-\eqref{eq:eulereq} around a background state: we set $\rho=\rho_0+\epsilon\rho_1+O(\epsilon^2)$ and $\psi=\frac{\delta\rho}{\rho}$ and we neglect the quantum pressure (last term in \eqref{eq:eulereq}) to arrive at
\begin{equation}
    \label{eq:waveeqsound}
    \begin{split}
            -\partial_t 
        \left(
            \frac{\rho_0}{c_s^2}
                \left(
                    \partial_t\psi_1+\pmb{v_0}\cdot\pmb{\nabla}\psi_1
                \right)
            \right)&+\\
    \pmb{\nabla}\cdot
        \left(
            \rho_0\pmb{\nabla}\psi_1-\frac{\rho_0\pmb{v_0}}{c_s^2}
            \left(
               \partial_t\psi_1+\pmb{v_0}\cdot\pmb{\nabla}\psi_1
            \right)
        \right)&=0.
    \end{split}
\end{equation}
In this equation, $c_s\equiv\sqrt{\frac{\hbar g \rho_0}{m}}$ is the local speed of sound and $\pmb{v_0}=\{v_0^x,v_0^y\}$.

Eq.\eqref{eq:waveeqsound} is strictly isomorphic to the wave equation of a massless scalar field propagating on a 2+1D curved spacetime \footnote{In the covariant notation, the upper and lower Greek indices are indices of coordinates.
These indices are used for the time and space components (in this order, ranging over the indexing set $\{0,1,2\}$, equivalent to the traditional $\{t,x,y\}$). 
Repeated indices are automatically summed over: $y=\sum_{i=0}^2 c_ix^i=c_0x^0+c_1x^1+c_2x^2$ is simplified by convention to $y=c_ix^i$ (Einstein summation convention). $\eta_{\mu\nu}$ is the inverse of $\eta^{\mu\nu}$ (contravariant metric).}:
\begin{equation}
    \label{eq:waveeqmlsf}
    \Delta\psi_1\equiv\frac{1}{\sqrt{-\eta}}\partial_\mu\left(\sqrt{-\eta}\eta^{\mu\nu}\partial_\nu\psi_1\right)=0.
\end{equation}
Being the spacetime three dimensional, the metric tensor $\eta$ is written in terms of a $3\times3$ symmetric matrix with $\eta=\mathrm{det}(\eta_{\mu\nu})$, and
\begin{equation}
    \label{eq:metric}
    \eta_{\mu\nu}=\frac{\rho_0}{c_s^2}\begin{pmatrix}
-\left(c_s^2-\pmb{v_0}^2\right) & -v_0^x & -v_0^y\\
-v_0^x & 1 & 0\\
-v_0^y & 0 & 1
\end{pmatrix}.
\end{equation}

The various components of this `acoustic metric' \cite{unruh_experimental_1981} are given by the fluid velocity.
Depending on the configuration of the flow, \textit{i.e.}, on $v_0^x$ and $v_0^y$, one can set up different fluid flows and study the propagation of fields on different space-times such as: (i) an accelerating flow along one spatial dimension only, (ii) a fluid flowing at increasing radial velocity in two spatial dimensions, (iii) a rotating flow like a vortex.
These various flow configurations may feature different regions characteristic of a black hole spacetime which may all be read from the metric \eqref{eq:metric}.
Notably, there is an event horizon in all three configurations --- where the flow velocity equals the speed of sound in (i)\footnote{In the experimental configuration of Fig.\ref{fig:2dflow}, this would be where $\sqrt{v_0^x+v_0^y}=c_s$.} and where the radial component of the fluid velocity equals the speed of sound ($v_r=c_s$) in (ii) and (iii).
Configuration (iii) also has an ergoregion, which is bounded by the surface where the flow velocity exceeds the speed of sound ($v_0^2>c_s^2$) \cite{visser_acoustic_1998}.

Polaritons can be optically tailored to generate either of these configurations.
Take the event horizon: when the polaritons are excited with a small wavenumber $k$ and a large enough intensity, the flow is superfluid inside the excitation area.
Outside the excitation area, the polariton density $\rho_0$ decreases, due to the dissipative nature of the polaritons, and therefore the interaction energy decreases.
This induces a reduction of the speed of sound ($c_s\propto\sqrt{\hbar g\rho_0}$).
Moreover, due to the conservation of the polariton energy in the cavity, the sum of the kinetic energy and of the interaction energy is constant \cite{Amelio}:
\begin{equation}
    \label{eq:energyconser}
    \frac{mv_0^2}{2}+\hbar g\rho_0=\Delta,
\end{equation}
where $v_0$ is the polariton flow velocity and $\Delta$ is the energy detuning at low power between the pump laser and the polariton branch at zero velocity.
This induces an increase of the kinetic energy and then of the flow velocity.
These effects result in the change of the flow characteristics from superfluid/subsonic to supersonic thus creating a horizon. As shown in Refs.  \cite{carusotto_fluidlightproposal_2012,Amelio}, the sharpness of this transition depends significantly on the shape of the pump spot and on the presence of an obstacle.

\section{One- and two-dimensional horizons}

\begin{figure*}[ht]
    \centering
    \includegraphics[width=.8\textwidth]{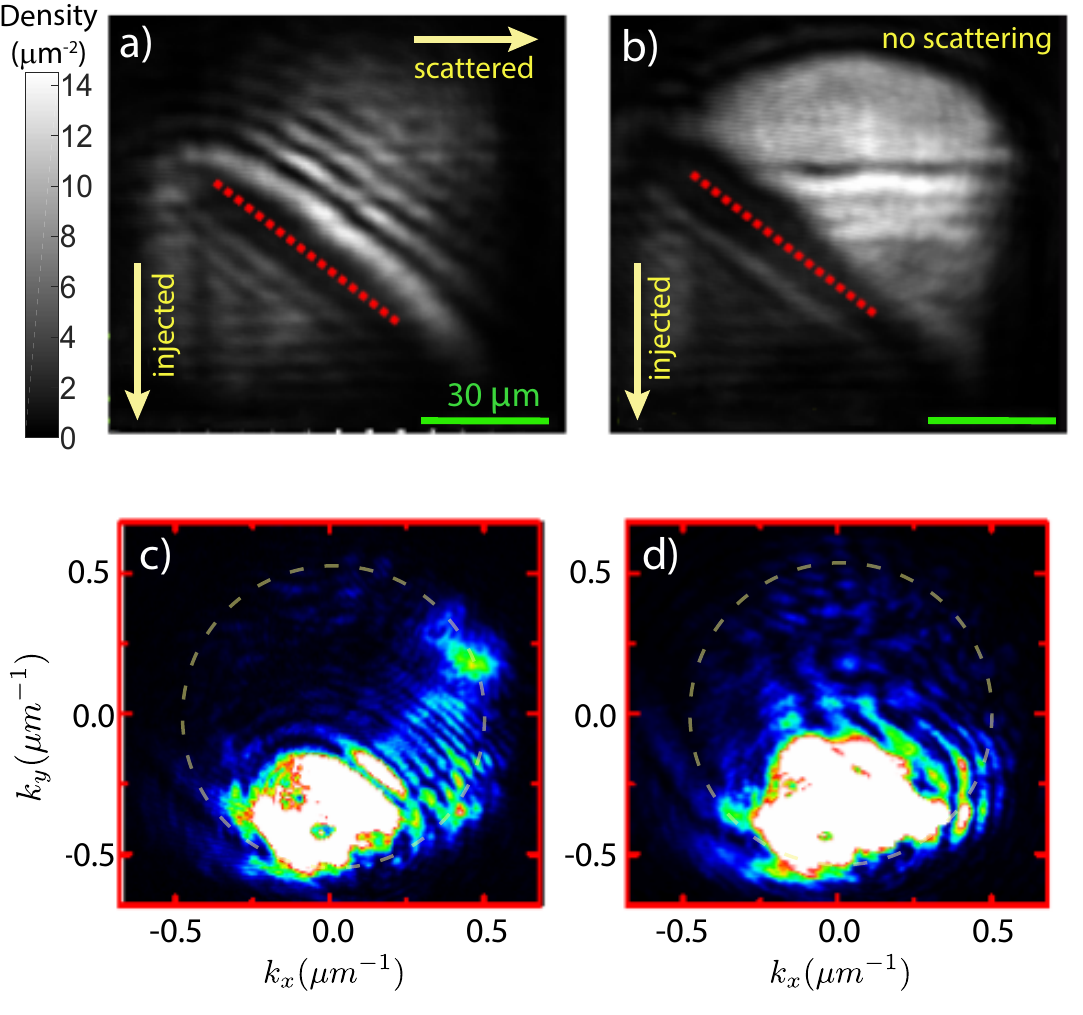}
    \caption{One-dimensional horizon in a polariton flow. Images of the propagating flow of polariton in downward direction as indicated by the yellow arrow in the near field (density, upper images) and in the far field (momentum, lower images); \textbf{a)} and \textbf{c)} show the low density case when the flow hits a barrier (optically generated defect) represented by the red dotted line; reflection on the barrier gives rise to interference fringes in the near field and to scattered light in the far field; \textbf{b)} and \textbf{d)} show the high density case with superfluidity which suppresses scattered light, with the disappearance of the fringes. The flow on the other side of the barrier is supersonic.    \label{fig:2dflow}}
\end{figure*}

In order to improve the sharpness of the horizon a defect can be placed close to the border of the pumping zone, as proposed in \cite{carusotto_fluidlightproposal_2012} and demonstrated by etching \cite{nguyen_acoustic_2015} a defect in the cavity.
Here we demonstrate the use of a defect generated optically, see Fig.\ref{fig:2dflow}.
The defect is generated by a control laser beam  with orthogonal relative polarisation to avoid interferences.
The interaction between polaritons with orthogonal polarisations is weaker than the interaction between polaritons with the same polarisation, but this can be compensated by increasing the power of the control beam creating the obstacle.
Here the linear obstacle (not directly visible, represented by a red dotted line) is placed at the border of the pumped region. The defect is at a 45$\degree$ angle to the polariton flow, which has a velocity of 0.64 $\mathrm{\mu m/ps}$.

In the low density regime, one can observe interference fringes between the incoming and reflected flows.
At high density, since the fluid velocity is lower than the speed of sound ($c_s = 0.8\, \mathrm{\mu m/ps}$), the system is in the superfluid regime.
In this case, there is no reflected flow, as attested by the disappearances of the interference fringes in the density distribution and of the reflected light in the momentum distribution.
Due to the driven dissipative nature of the polariton system linked to the cavity losses, part of the population in the superfluid pumped region goes out of the cavity. 
The remaining part of the flow goes through the barrier, and in this unpumped region, the flow has a much lower density.
The density decreases by about a factor of 2 and the speed of sound is now $0.56\,  \mathrm{\mu m/ps}$.
Moreover, due to the energy conservation the fluid velocity increases and is now $0.90\, \mathrm{\mu m/ps}$, enhancing the supersonic character of the fluid \cite{adrados_phd_2011}.
This experiment demonstrates the presence of an event horizon, with a sharp transition between the upstream (subsonic) and downstream (supersonic) regions.
The spatial mode of the laser fixes the derivative of the flow velocity at the horizon. Here, a Gaussian beam was used but it is possible to create steeper (and sharper) horizons by shaping the laser beam with \textit{e.g.} a spatial light modulator.

\begin{figure*}[ht]
    \centering
    \includegraphics[width=.65\textwidth]{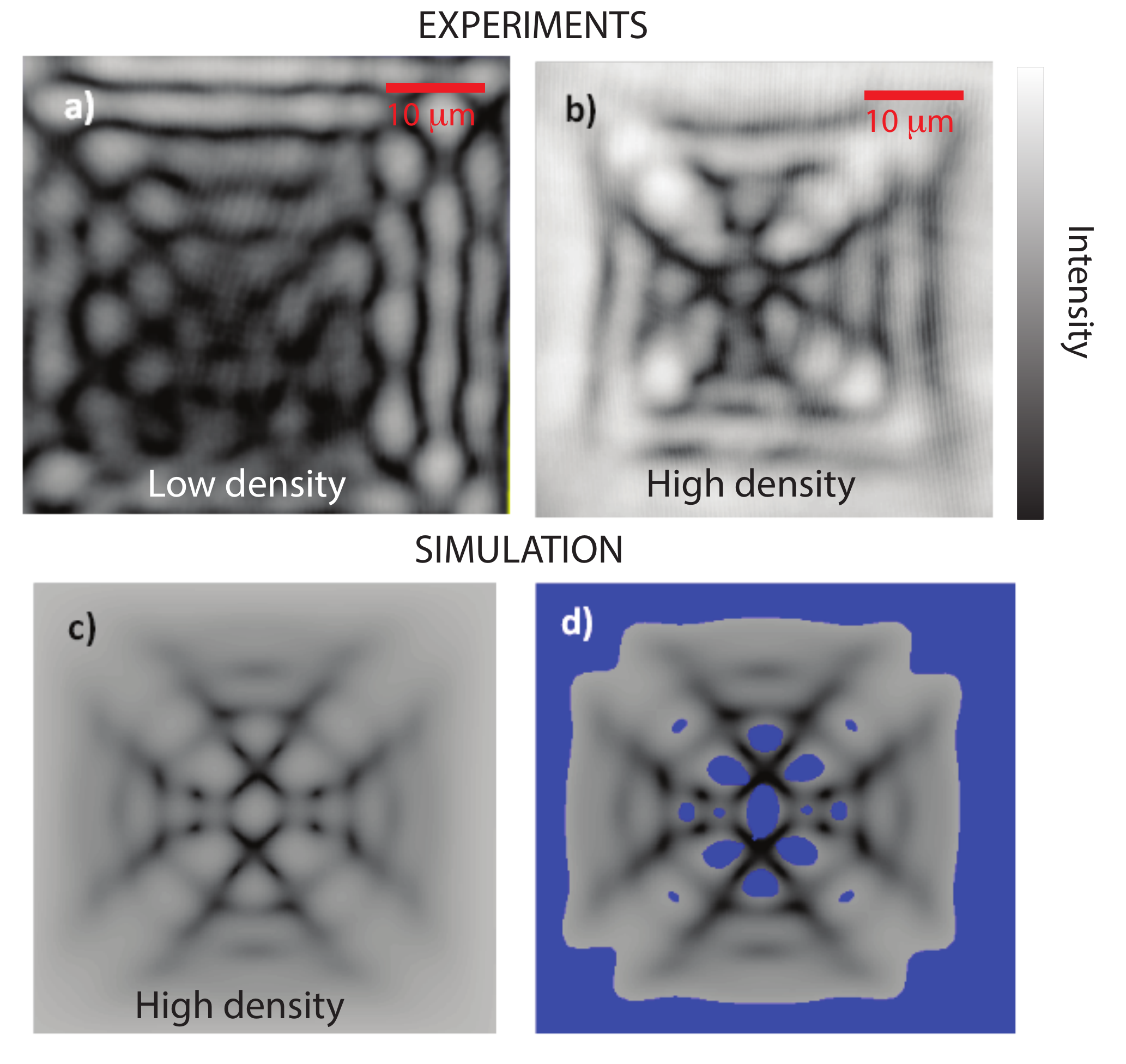}
    \caption{Two-dimensional sonic horizon. Image of converging polariton flows, starting from the sides of the square; \textbf{a)} low density regime, with interference fringes; \textbf{b)} high density regime, with less interference fringes, especially near the edges; \textbf{c)} Simulation of \textbf{b)}. \textbf{d)} Velocity map at high density. Regions in blue are subsonic, the grey region is supersonic.
    \label{fig:mach}}
\end{figure*}

This process can be extended to create a closed 2D horizon.
By generating a polariton ensemble arranged in a square, one produces a converging flow towards the centre of the square. For this a pump laser beam at $k = 0$ is focused on the microcavity surface.
A square mask is placed in the pump laser beam, and is imaged on the microcavity.
The image of the mask generates a square dark region (with a  $45\,\mathrm{\mu m}$ side) at the centre of a bright Gaussian spot (with diameter at half maximum of $100\,\mathrm{\mu m}$).
At low intensity (Fig.\ref{fig:mach} \textbf{a)} polaritons flow from the four sides of the mask towards the centre, generating four plane waves that create an interference pattern clearly visible on the figure.
Vortex-antivortex pairs also form because of the local disorder.
At high intensity (Fig.\ref{fig:mach} \textbf{b)}, the polaritons are generated resonantly at k=0 outside of the masked region and then enter by diffraction in the masked region, where they acquire a momentum in order to conserve energy.
The repulsion between polaritons leads to an enlargement of the lattice unit cell due to an enhancement of the interaction, as confirmed by the numerical model (Fig.\ref{fig:mach} \textbf{c)} and to the appearance of a superfluid region.

From this image the Mach-number ($M=\frac{v}{c_s}$) is calculated using the local speed of sound of the polariton ensemble (Fig.\ref{fig:mach} \textbf{d)}).
In the region inside the trap that is close to the borders, the system is in a subsonic/superfluid regime due to its high density, which is consistent with the disappearance of interference fringes.
As polaritons move towards the centre of the trap, their density decreases (this is due to their finite lifetime) and so does the sound velocity of the fluid --- the fluid becomes mainly supersonic (except at the centre, where the four flows merge, creating an inner subsonic region).
There is an event horizon between the subsonic and supersonic regimes in this convergent flow near the edge of the square.
The velocity and the density of the polariton flow can be adjusted to modify the parameters of this horizon. 

It has been theoretically predicted that it is possible to observe the Hawking effect at these sonic horizons \cite{nguyen_acoustic_2015,carusotto_stimulatedfluid_2016}.
We propose to begin by seeding the scattering process with a classical, low amplitude excitation (similar to \cite{euve_scattering_2018}).
This can be done with a probe beam with an energy higher than that of the polariton flow, propagating in the downstream supersonic region towards the horizon. 
Such a probe wavepacket can be partly transmitted and reflected at the horizon: we will thus retrieve the coefficient of the scattering matrix that rules the Hawking effect \cite{Macher_blackwhite_2009,Larre_scattering_2012,Jacquet_quantum_2015,jacquet_analytical_2019} and observe the Bogoliubov branches of the dispersion by using a method recently implemented in paraxial fluids of light in hot atomic vapours \cite{fontaine_disprel_2018}.

The spontaneous Hawking process results from the conversion of vacuum fluctuations into correlated phonons on the two sides of the horizon \cite{boiron_quantum_2015}, which can be detected as intensity correlations in the near-field emission of the cavity \cite{Balbinot_correlations_2008,Carusotto_numerical_2008,isoard_quantum_2019}.
As shown in \cite{carusotto_stimulatedfluid_2016}, these correlations correspond to very specific branches of the Bogoliubov dispersion on each side of the horizon, one of them being the negative fluid-frame frequency branch.
Their detection, achieved only in cold atoms \cite{munoz_de_nova_observation_2019}, is non trivial and remains a challenge for polariton fluids, as a detailed theory for the correlation signal --- that would account for the thermal noise (including stimulation of the Hawking effect at the horizon by the ever-present non-thermal, incoherent bath of phonons in the system \cite{Busch_entanglementfluid_2014}) --- is still missing.
Yet, quantum noise in semiconductor microcavities was studied experimentally using homodyne detection \cite{boulier_squeezing_2014}.
This method offers a high sensitivity and should be very efficient for studying the Hawking effect.
Further, a measure of the ``quantumness of the output state'' would discriminate correlations resulting from spontaneous emission from those resulting from noise (although additional noise may degrade the signal to noise ratio in the measurement of Hawking correlations).
To this end, one could use various measures such as the Perez-Horodecki criterion \cite{campo_inflationary_2006,nova_cauchyschwarz_2014,Busch_entanglementHR_2014,Busch_entanglementfluid_2014,finazzi_entangled_2014,boiron_quantum_2015,coutant_low-frequency_Gennes_2018,coutant_low-frequency_Vries_2018}, the logarithmic negativity \cite{jacquet2020influence} or squeezing  \cite{boulier_squeezing_2014} to determine whether the output state is nonseparable.
Importantly, entanglement monotones such as the logarithmic negativity would provide unprecedented insight in the quantum statistics of Hawking radiation.

\section{Rotating geometry}
\begin{figure*}[ht]
    \centering
    \includegraphics[width=.7\textwidth]{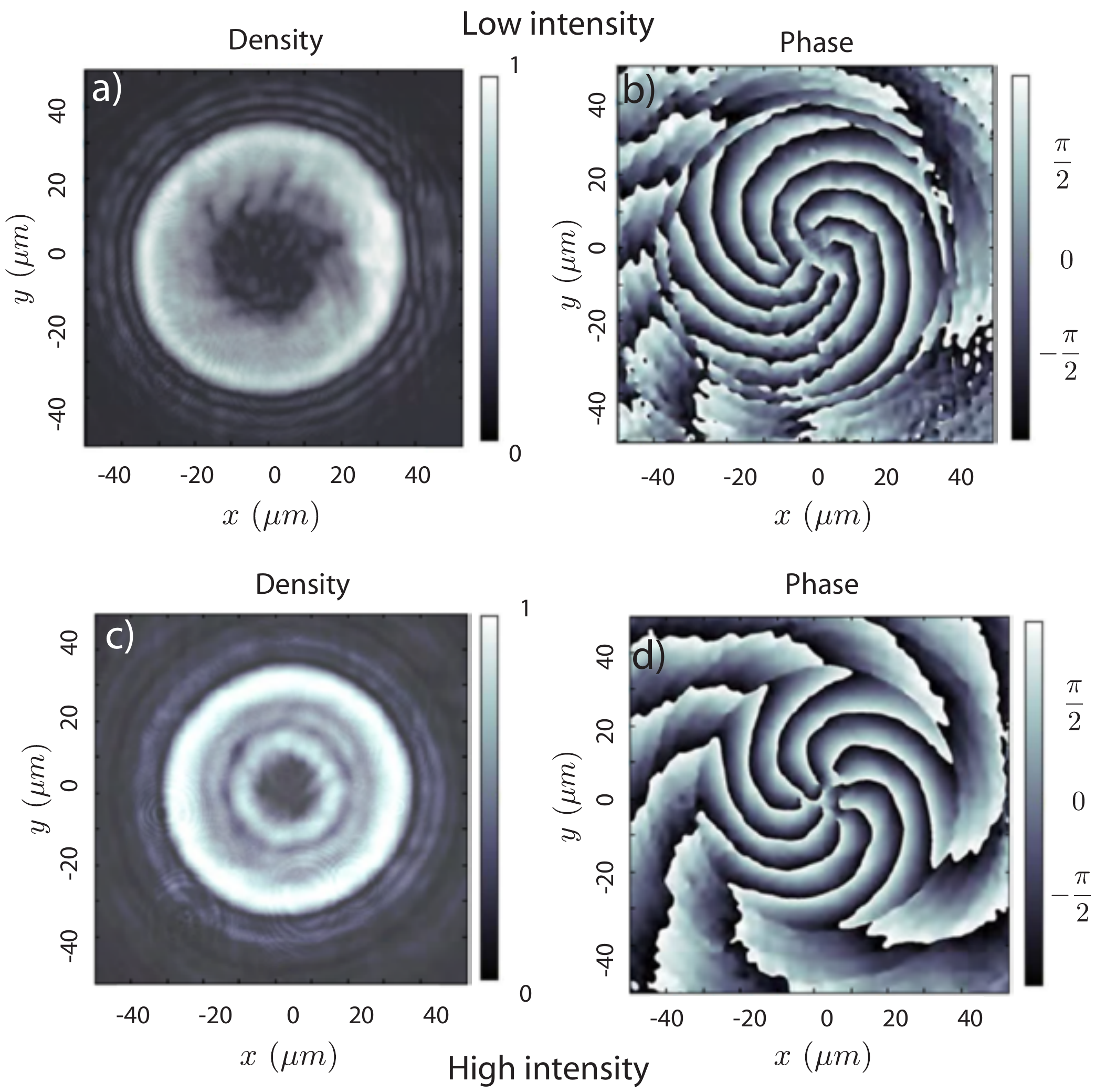}
    \caption{Rotating geometry. Density (left panels) and phase (right panels) maps of the polariton system excited with a Laguerre-Gauss beam with l=8 for low and high density regimes. Black hole surfaces: ergosurface, green dotted line; event horizon, red dotted line.
    \label{fig:rotatingflow}}
\end{figure*}

In the microcavity driven-dissipative system, the use of a pump laser to generate the polaritons also allows to inject an orbital angular momentum in the fluid. 
This can be achieved by using a Laguerre-Gauss (LG) beam produced by a phaseplate or a spatial light modulator (SLM) \cite{Boulier2015,Vocke_rotating_2018}.
Fig.\ref{fig:rotatingflow} shows the density and phase map obtained with a beam with $l=8$ for low and high density regimes.
Polaritons rotate and propagate towards the centre with an azimuthal and a radial velocity.
A low density region is present in the centre.
In the low density regime, the phase map shows some disorder.
In the high intensity regime, the disorder decreases due to the interactions and, in the density map, one can observe a concentric ring-like structure at the centre of which a dark zone similar to the low density case is visible \cite{bouli2014}. 

The flow configuration of Fig.\ref{fig:rotatingflow} is not only rotational: there is an influx of polaritons towards the centre of the beam, where they naturally escape from the cavity. 
(Here, polaritons losses are not localised solely in the center of the beam but occur everywhere in the cavity.)
The ability to create a rotating flow with a drain is a key advantage of lossy systems such as polaritons in microcavities over conservative systems such as atomic ensembles for the simulation of a rotating black hole.
For example, ref. \cite{Solnyshkov_penrose_2019} shows that focusing a Laguerre-Gauss beam (with a maximal angular momentum $l=18$) with sufficient power on a microcavity creates an analogue rotating black hole with a diameter of the order of a 20 to 30 micrometres.
In this case, there is an ergoregion: a region limited on the outside by a static limit, where the total flow velocity is equal to the velocity of sound and on the inside by the inner horizon, where the radial flow velocity is equal to the velocity of sound.

The flow velocities can be measured in Fig.\ref{fig:rotatingflow} using the gradient of the fluid phase as presented in Section \ref{sec:anagrav}. 
Although the signal to noise ratio does not allow to reliably determine the position of the event horizon and of the ergosurface, the radial and total velocities are found to vary from $0.5\mu$m/ps outside the beam to over $2\mu$m/ps when going towards the center of the beam, while the maximum sound velocity is of the order of $1\mu$m/ps.
This shows that the system has adequate parameters to define an ergosphere and a horizon.
Polaritons are thus a promising system for generating a geometry in which the physics of rotating black holes can be studied.

Thus, in this system one can test the Penrose effect \cite{penrose_extraction_1971}, which is at the heart of instability-driven phenomena such as the black-hole bomb.
The Penrose effect involves the creation of a pair of positive and negative energy particles --- vortices and antivortices  in the present case \cite{Solnyshkov_penrose_2019}.
A pair of vortex-antivortex can be generated by creating a localised density dip in the ergosurface with a laser.
The antivortex corresponds to the negative energy particle and is absorbed inside the horizon, annihilating a unit of angular momentum.
The vortex can escape from the horizon through the static limit.
This system is also well adapted to observe the Zeld’ovich effect \cite{zeldovich_amplification_1972,Cardoso_detecting_2016,Prain_superradiance_2019} and its correspondent effect in rotating geometries: superradiance via the amplification of waves scattered by a rotating obstacle.
The observation of stimulated superradiance would also enable the study of other classical effects of rotating geometries, such as the apparition of light-rings \cite{torres_lightrings_2018}, the generation of quasi-normal modes \cite{patrick_black_2018,torres2018application,torres_analogue_2019} (damped oscillations emitted upon kicking the vortex) and even a classical effect of the rotating flow on the fluid akin to back-reaction \cite{goodhew2019backreaction}.
Further, observing spontaneous superradiance could lead to the study of both semi-classical and purely quantum effects of fields on rotating geometries, such as the rise of a grey-body factor and spontaneous back-reaction of the emitted quanta on the polariton metric.

\section{Conclusion}
In conclusion, polariton fluids in microcavities are a very promising system for analogue gravity.
In particular, we have shown how using an optical defect permits the generation of various flow profiles without engineering of the cavity.
We have demonstrated the creation of static one- and two-dimensional acoustic black hole horizons in polariton flows.
As for the latter, a more regular system could be created with an annular-shaped pumping beam.
Thanks to the rather high Hawking temperature predicted in these systems the detection of Hawking radiation should be possible from one- or two-dimensional horizons, thus enabling the first measure of entanglement in an analogue experiment.
We have also created a rotating flow implying an ergoregion and an inner horizon, which opens the door to the future observation of the stimulated and spontaneous Penrose and Zeld'ovich effects as well as of regimes of instabilities in black hole systems.
\vskip6pt

\enlargethispage{20pt}

\textbf{Acknowledgements}

The authors gratefully acknowledge discussions with I Carusotto and J Bloch.
We acknoweldge the support of ANR projects C-FLigHT (ANR-16-ACHN-0027) and QFL (ANR-16-CE30-0021). This work has received funding from the European Union's Horizon 2020 Research and Innovation programme under grant agreement No. 820392 (PhoQuS). A.B. and Q.G. are members of the Institut Universitaire de France (IUF).



\end{document}